\documentclass[12pt]{iopart}

\usepackage[english]{babel}
\usepackage[latin1]{inputenc}
\usepackage{amssymb}
\usepackage{mathrsfs}

\usepackage[dvips]{graphicx}
\usepackage{epsfig,bm}

\newcommand{\prt}{\partial}
\newcommand{\al}{\alpha}

\newcommand{\Dp}{\Delta p}
\newcommand{\Sv}{\bar{v}}
\newcommand{\Dv}{\Delta v}
\newcommand{\sinca}{s_{\alpha}}

\newcommand{\ca}{\cos{(\alpha)}}

\newcommand{\Th}{\Theta}

\begin{document}
\title[Short-time dynamics in presence of wave-particles interactions: a perturbative approach.]{Short-time dynamics in presence of wave-particles interactions: a perturbative approach.}
\author{R Bachelard}
\address{Centre de Physique Théorique\\
Campus de Luminy, Case 907\\
13288 Marseille cedex 9, France}
\ead{bachelard@cpt.univ-mrs.fr}

\author{D Fanelli}
\address{Theoretical Physics Group\\
  School of Physics and Astronomy\\
  The University of Manchester\\
  Manchester, M13 9PL, UK}
\ead{duccio.fanelli@manchester.ac.uk }

\begin{abstract}
The self-consistent interaction between a beam of charged particles and a wave is considered, within a Vlasov picture.
The model is  discussed with reference to the case of a Free Electron Laser. Starting with
a spatially bunched {\it waterbag} distribution, we derive, via perturbative methods,   
closed analytical expressions for the time evolution of the main macroscopic observables. Predictions of the theory 
are shown to agree with direct numerical simulations. 
\end{abstract}

\pacs{52.65.Ff, 41.60.Cr}

\maketitle

\section{Introduction}

Wave-particles interaction is a basic process in physics, which  is 
encountered in a large class of different phenomena. Most spectacular examples 
are undoubtely found in astrophysical context, but also in 
laboratory plasmas where technological aspects of nuclear fusion are adressed.

Free-Electron Lasers (FELs) \cite{sase,seed} 
constitute another important application where the dynamical 
interplay between particles and waves is well known to be central.  
The physical mechanism that drives the process 
of light amplifiction is in fact the interaction  between a relativistic electron
beam and a co-propagating optical wave, inside the so--called {\it undulator}. The latter  
generates  a magnetostatic periodic field, thus forcing the electrons to follow curved trajectories and emit 
synchrotron radiation. This incoherent light seeds, termed spontaneous emission, cumulate under resonance condition, 
and eventually result in the coherent laser signal. In a single--pass scheme, the laser is produced during a single passage 
inside the undulator, and the basic features of the system dynamics are successfully captured by a simple one-dimensional 
Hamiltonian model \cite{Bonifacio}. 

Remarkably, an analogous mathematical formulation is shown to describe   
the electron beam-plasma instability, a basic problem in   
kinetic plasma turbulence \cite{ElskensBook}.
When a weak electron beam is injected into a thermal plasma,
electrostatic modes at the plasma frequency (Langmuir modes) are
destabilized and, subsequently, amplified \footnote{Travelling Waves Tubes (TWTs) \cite{TWT} 
have been extensively adopted to mimic beam-plasma interactions. The amplification process in a TWT can in fact be
described in term of an analogous Hamiltonian setting.}. 


In both cases, the field intensity experiences a sudden growth, which is followed by a 
non linear saturation towards a non-equilibrium plateau. This initial violent relaxation is governed by the 
the Vlasov equation, a partial differential equation which represents the continuous counterpart of the 
discrete dynamics. Based on the Vlasov description, one can quantitatively predict the average behaviour 
of the system at saturation \cite{julien} and derive a reduced formulation to track the detailed time evolution of the 
main collective variables \cite{tennyson,andrea1,andrea2}.
According to this latter scenario, a significant number of particles experience a bouncing motion 
in one of the (periodically repeated) potential wells, 
and form a clump that evolves as a single macro-particle localized in space. 
The remaining particles populate the surrounding halo, being almost
uniformly distributed in phase space between two sharp boundaries. 

In real devices, however the finite extension of the interaction region, e.g. the size of the undulator, 
prevents the system to enter the deep saturated state and the initial sudden growth is the only regime that can be 
practically explored. It is therefore of general interest to mathematically address the study of the system dynamics for 
short times, aiming at providing closed analytical expressions that apply to a large class of initial condition. 

When initializing the system around an equilibrium condition, one can straightforwardly linearise the equations of motion and 
obtain an estimate that is shown to accurately agree with the numerics. However, non-equilibrium initial conditions are also experimentally relevant  
\cite{FELbunched} and result in a substantially different dynamics. In this paper, we shall focus on this latter 
case and derive perturbative solutions for the broad class of spatially non-homogeneous water bag initial profiles.     
 
The paper is organized as follows~: In Section \ref{model} we
introduce the one-dimensional model of a FEL amplifier
\cite{Bonifacio}. The continuum limit is also discussed and the Vlasov model 
presented. Section  \ref{derivation} is devoted to presenting key assumptions of the perturbative calculations. 
Closed expressions for the main macroscopic quantities are derived in Section \ref{macroscopic} and compared with
numerical simulations. Finally, in Section \ref{conclusion} we sum up and draw our conclusions.

\section{From the Hamiltonian model to the Vlasov equations: the case of the FEL}\label{model}

In the following we shall focus on the case of a Free Electron Laser. However, as previously noticed,
the model is more general, and can be regarded as a paradigmatic example of systems with wave-particles interactions.
In this respect, the conclusions of our analysis will apply to other physical contexts 
where the same basic mechanism holds. 

Under the hypothesis of one-dimensional motion
and monochromatic radiation, the steady state 
dynamics of a Single-Pass Free Electron Laser is described by the
following set of equations:

\begin{eqnarray}
\frac{{\mathrm d}\theta_j}{{\mathrm d}\bar{z}} &=& p_j\quad, \label{eq:mvttheta}\\
\frac{{\mathrm d}p_j}{{\mathrm d}\bar{z}} &=&
-Ae^{i\theta_j}-A^{\ast}e^{-i\theta_j}\quad,
\label{eq:mvtp}\\
\frac{{\mathrm d} \bm{A}}{{\mathrm d}\bar{z}} &=&
 \frac{1}{N} \sum_j e^{-i\theta_j}~\quad, \label{eq:mvtA}
\end{eqnarray}

where $\bm{A}=A_x+i A_y$ represents the wave vector potential, $\theta_j$ stands for the phase of the electron $j$ with respect to the ponderomotive wave, while $p_j$ is its rescaled energy. 
All are adimensional quantities and the reader can refer to \cite{Bonifacio} for a detailed account on the derivation of the 
model and an exhaustive connection with the physical parameters of the machine. We shall here simply recall that 
$\bar{z}$ is the rescaled longitudinal coordinate, inside the undulator, which essentially plays the role of time. In the following, for the sake of simplicity, 
we shall replace it by $t$. 

The above system of equations ($N$ being the number of electrons) can be derived from the Hamiltonian 
\footnote{In the following we set the detuning parameter to zero, thus assuming perfect resonance condition. The analysis
can be however extended to the case where an energy mismatch has to be accounted for.}:
\begin{equation}
H=\sum_{j=1}^N\frac{p_j^2}{2} + 2\sqrt{\frac{I}{N}}\sum_{j=1}^N
\sin(\theta_j-\phi) \label{eq:Hamiltonian},
\end{equation}
where the intensity~$I$ and the
phase~$\phi$ of the wave are given by
$\bm{A}=\sqrt{I/N} \exp(-i \phi)$. Here the canonically conjugated variables are
$(p_j,\theta_j)$ for $1 \leq j \leq N$ and $(I,\phi)$. 
Besides the ``energy'' $H$, the total momentum $P=\sum_j
p_j +  I$ is also conserved.

In the continuum limit Eqs.(\ref{eq:mvttheta})--(\ref{eq:mvtA}) are rigorously mapped into the following system of partial differential equations:

\begin{equation}\label{df} \frac{\prt f}{\prt t} = p \frac{\prt f}{\prt \theta} - 2 \left( A_x \cos \theta -A_y \sin \theta \right)
\frac{\prt f}{\prt p},
\end{equation}
\begin{equation}\label{dax} \frac{\prt A_x}{\prt t}  = \int_{-\pi}^{\pi} d\theta \int_{-\infty}^{\infty} dp f(\theta,p,t) \cos \theta,
\end{equation}
\begin{equation}\label{day} \frac{\prt A_y}{\prt t}  = - \int_{-\pi}^{\pi} d\theta \int_{-\infty}^{\infty} dp f(\theta,p,t) \sin \theta,
\end{equation}

where $f(\theta,p)$ represents the single particle distribution. The bunching coefficients $b_k$ are in turn defined as:

\begin{equation}\label{bunch} b_k(t) = \int_{-\pi}^{\pi} d\theta \int_{-\infty}^{\infty} d p \exp(-i k \theta) f(\theta, p, t) \qquad for \quad k=1,2,3.. 
\end{equation} 

and measure the degree of spatial packing of the particles.
 
Assuming periodic boundary condition, the system (\ref{df})-(\ref{day}) admits the following stationary solution:

\begin{equation}\label{soleq} A_x=A_y =0 \qquad f=f_0(p)
\end{equation} 

Linearizing around equilibrium one can derive an approximate solution that holds for relatively short times. To this end, the following ansatz is put 
forward: 

\begin{equation}\label{pertubrsolu}
  f(\theta,p,t)=f_0(p)+f_1(\theta,p,t),\quad A_x(t)=X_1(t)\quad\mbox{and}\quad A_y(t)=Y_1(t)\quad.
\end{equation}

Introducing in system~(\ref{df}) and using the notation  
$\eta(p)=\partial_p f_0$, we obtain at lowest order

\begin{eqnarray}
 (\partial_t+p\partial_\theta)f_1&+\eta(X_1 \cos \theta&-Y_1 \sin \theta)=0\label{deuxsolia}\\
 \int_{-\pi}^{\pi}\!\! d \theta\!\!\int_{-\infty}^{+\infty}\!\! d p\  f_1\cos \theta&
 -\frac{d X_1}{d t}&=0\label{deuxsolib}\\
  \int_{-\pi}^{\pi}\!\! d \theta\!\!\int_{-\infty}^{+\infty}\!\! d p\  f_1\sin \theta&
 +\frac{d Y_1}{d t}&=0\label{deuxsolic}
 \end{eqnarray}

Such a linear system admits the following normal modes solution
\begin{eqnarray}
f_1(\theta,p,t)&=&F_1(p)\, e^{i(\theta-\omega t)}+F_1^*(p)\, e^{-i(\theta-\omega^* t)}\label{solf1}\\
X_1(t)&=&X_1\, e^{-i\omega t}+X_1^*\, e^{i\omega^* t}\\
Y_1(t)&=&iY_1 \, e^{-i\omega t}-iY_1^*\, e^{i\omega^*
t}\quad.\label{solY1}
\end{eqnarray}
where the symbols $*$ stands for the complex conjugate and $\omega\in \mathbb{C}$. By introducing solution 
(\ref{solf1})-(\ref{solY1}) into the linearized system one obtains, after some algebra, the following dispersion relation:

\begin{equation}\label{dispersionfinal}
\omega =  \int_{-\infty}^{+\infty}\!\! dp
\,\frac{\eta(p) }{p-\omega}\quad
\end{equation}

which can be solved with respect to $\omega$, once the equilibrium initial condition $f_0(p)$ is specified. If 
a complex solution exists, the field grows exponentially, see Eqs. (\ref{solf1})-(\ref{solY1})), otherwise it oscillates 
indefinitely (see also \cite{Bonifacio}). 

It should be however stressed that the above treatment applies if the system is locally pertubed around the equilibrium initial condition
(\ref{soleq}). For more general out-of-equilibrium settings, the linearization fails and other 
strategies need to be developed. In particular, we shall here discuss a perturbative approach aimed at characterizing 
the evolution of the system initialized in the so--called water--bag state, thus allowing for a spatial bunching ($b_k(0) \ne 0$) of the beam. 
This technique enables us to derive closed analytical expressions for the time evolution of the  
fundamental macroscopic observables that characterize the system dynamics..


\section{Simplified water-bag approximations}
\label{derivation}

In the following we shall consider an initial water--bag profile: particles are confined in a finite portion of phase--space and there display a 
uniform distribution. This is a rather common choice already invoked in several studies \cite{julien} and often assumed to provide a simplified 
description of more realistic initial conditions.

Our perturbative analysis is based on a simple assumption: we hypothesize that the initial particles' evolution changes the shape of the water-bag, 
while preserving the uniform distribution (homogeneous density) inside the stretched domain. This working ansatz is corroborated by direct numerical inspection and shown to hold
approximately during the initial violent relaxation, until the saturated regime is eventualy attained. From there on, a dense cluster starts to 
develop and one has to resort to the so-called macro-particle scheme (see \cite{tennyson, andrea1, andrea2}) to derive a reduced theoretical framework.

\begin{figure}[ht]
\centerline{
$\begin{array}{cc}
\epsfig{figure=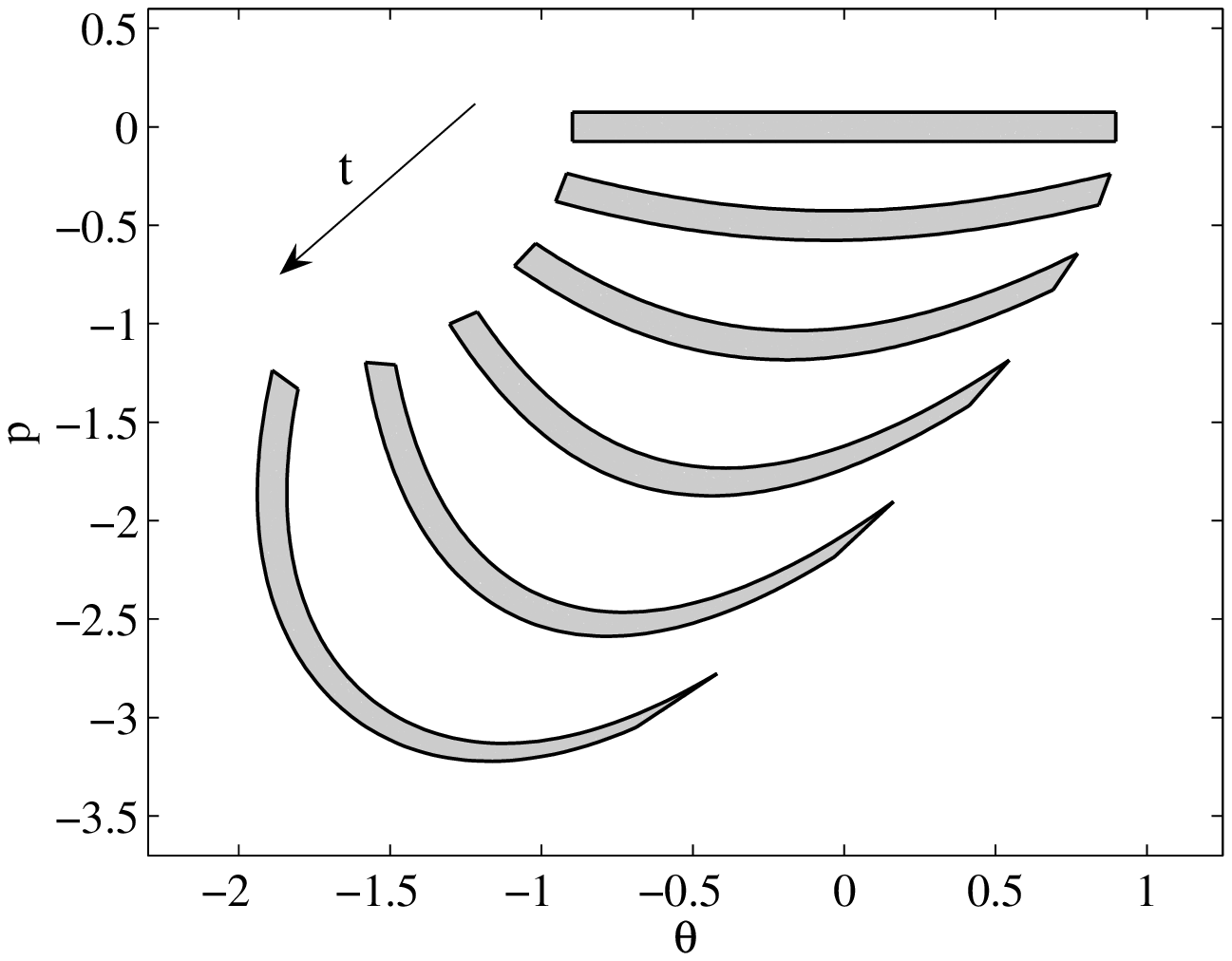,width=10cm} & \epsfig{figure=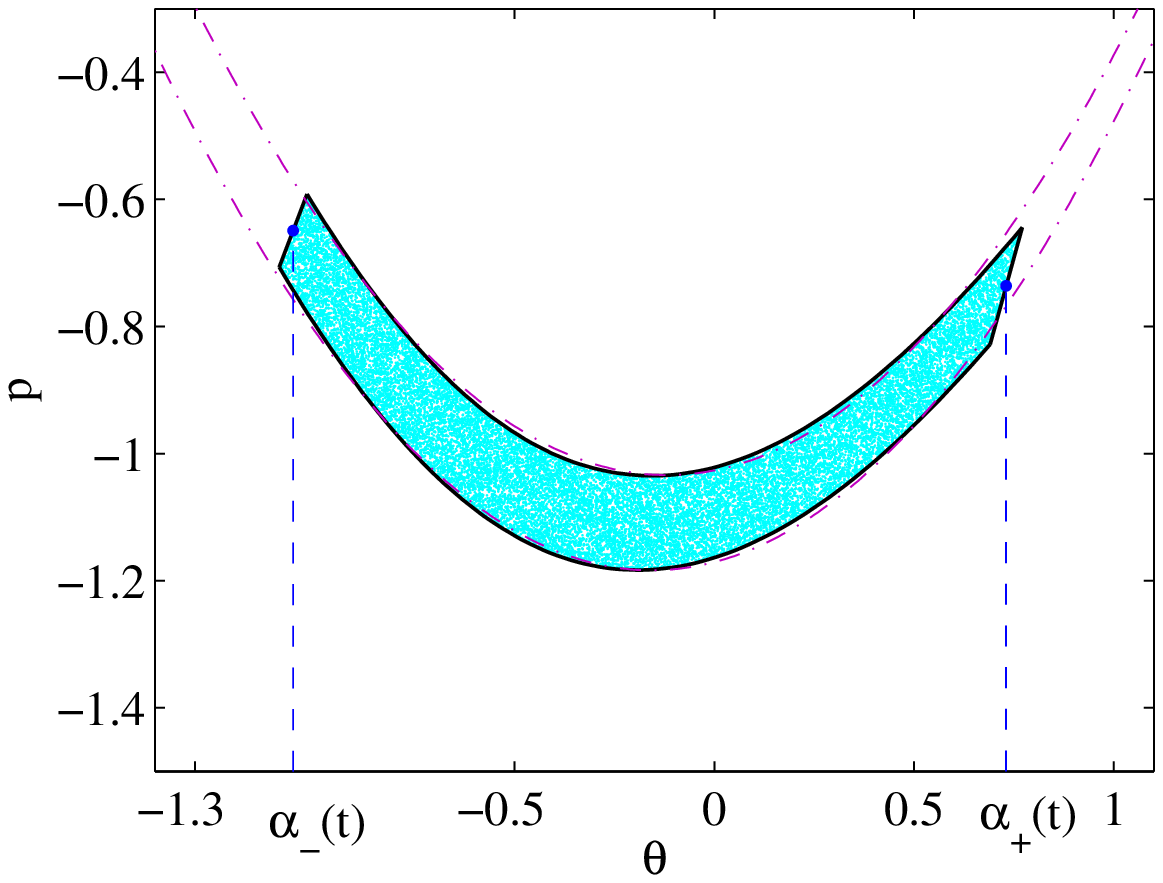,width=10cm}
\end{array}$}
\caption{Left~: Evolution of the bunch of particles in phase-space, at $t=0,\ 0.25,\ 0.5,\ 0.75,\ 1,\ 1.25,\ 1.5$ 
(from top right to bottom left). The boundaries $P_\pm$ are plotted in dark (horizontal boundaries), while the inner part of the waterbag 
stands in grey. At $t \approx 1.4$, the bunch flips~: this is clearly a limitation (in time) of our modelization of the system. 
Right~: waterbag at $t=0.5$; the dash-dotted lines correspond to the second degree polynomial fit of Eq.(\ref{ansppm}).\label{figPS}}
\end{figure}

According to the proposed formulation, and recalling that Liouville theorem holds, one can formally trace the water-bag evolution in term of its outer
boundaries, once the initial density has been assigned. Figure \ref{figPS} illustrates this concept pictorially, for the case of a rectangular 
water--bag domain which will be assumed in this study. Hereafter, the
upper and lower boundaries are parametrized as $P_+(\theta,t)$ and $P_-(\theta,t)$, a functional dependence which can be correctly invoked as long as the 
evolution stays single-stream, i.e. before a lateral flip occurs. As for the left (resp. right) lateral edge, we assume $\theta = \Th_-(t)$ 
(resp. $\theta = \Th_+(t)$), which in turn amounts to model it as a vertical, though dynamic, barrier, hence neglecting its inclination. These prescriptions 
translate in the following mathematical expression for the initial single-particle distribution function $f(\theta,p,0)$:

 \begin{equation}\fl f(\theta,p,0) = f_0 [\Omega(\theta-\Th_-(t))-\Omega(\theta-\Th_+(t))] [\Omega(p-P_-(\theta,t))-\Omega(p-P_+(\theta,t)],\end{equation}
 
 where $f_0$ labels the water-bag's density and $\Omega$ represents the Heavyside function. Moreover, we shall limit the discussion to initially symmetric 
 profiles which mathematically yields to:
 
\begin{equation} \Th_+(0)=-\Th_-(0)=\al \ \mathrm{ and } \ P_+(\theta,0) = -P_-(\theta,0)=\frac{\Dp}{2}, \forall \theta. \end{equation} 

Notice that for $\Th_+=-\Th_-=\pi$, one formally recovers the
 setting (\ref{soleq}) and an exponential growth is thus expected, provided $A_x=A_y=0$. 
In the following we will concentrate on the case $\alpha < \pi$. The normalization condition results in $f_0=1/ 2 \alpha \Delta p$.

Particles and wave are in phase at $t=0$, i.e. $\phi(0)=\Th_+(0)-\Th_-(0)$, a condition that results in the optimal growth of the field intensity initially set 
to $I_0$ (either finite or zero). Finally, we shall assume that the $P_+$ and $P_-$ profiles are accurately interpolated by two parabolas 
centered in $\theta=0$, and consequently parametrized as:

\begin{equation}\label{ansppm} P_{\pm} (\theta,t) = u (t) \theta^2 + v_{\pm} (t),
\end{equation}

with $v_\pm(0) = \pm \frac{\Dp}{2}$ and $u(0)=0$. The adequacy of this approximation is verified numerically via direct fit and further confirmed by a posteriori
testing the predictive ability of the self-consistent theory here developed. The linear term in $\theta$  does not appear in Eq.(\ref{ansppm}) as the 
associated coefficient scales as $t^3$, thus falling beyond the  accuracy of our perturbative scheme.

\section{Solving the short-time system dynamics: A perturbative derivation}

Starting from this initial setting and recalling Eq.(\ref{dax}), one straightforwardly obtains: 

\begin{equation} \dot{A_x} = f_0 (v_+ - v_-) (\sin{\Th_+} - \sin{\Th_-}) = \sinca+O(t),
\end{equation}

where $\sinca=\sin(\al)/ \al$. Integrating yields~:

\begin{equation}\label{ax} A_x(t)=A_x^0+\sinca t+ O(t^2),
\end{equation}

with $A_x^0=\sqrt{I_0}$. As for $A_y$, combining Eqs.~(\ref{day}) and (\ref{ansppm}) results in~:

\begin{equation}\label{day2} \dot{A_y} = f_0 (v_+ - v_-) (\cos{\Th_+} - \cos{\Th_-}) = O(t),
\end{equation}

since $\Th_+(0)=-\Th_-(0)$. This in turn implies 
\begin{equation}\label{ay} A_y(t)=O(t^2).
\end{equation}
 
We now look at the evolution of a reference particle (here baptised test-particle) 
located at the boundaries of the waterbag. The motion of such a particle is governed by the following  
Hamilton equations~:

\begin{equation}\label{hampart} \left\{ \begin{array}{lll} \dot{\theta}&=&p \\
 \dot{p} &=& -2(A_x \cos{\theta} - A_y \sin{\theta}) \end{array} \right. \end{equation}

Consider in particular a (virtual) particle of initial coordinates $(\theta(0)=\pm\al, p(0)=0)$. During the time interval covered by our investigations,
namely before the outer contour enters a multi-stream regime, it can be reasonably assumed that the particle evolves 
coherently with the associated boundary (see Fig.\ref{figPS}), an observation which suggests identifying $\theta(t)=\Th_\pm (t)$. 
According to Eqs.(\ref{ax})-(\ref{hampart}), the particle's position obeys to~:

\begin{eqnarray}\label{ccl} \ddot{\Th_\pm} & = & -2(A_x \cos{\Th_\pm} - A_y \sin{\Th_\pm}) \\
 & \simeq & -2 A_x^0 \cos{\al} + O(t). \nonumber \end{eqnarray}


The preceding equation can be integrated and results into $\Th_\pm(t) = \pm \al - A_x^0 \ca t^2 +O(t^3)$. This latter expression is then re--inserted into the first of 
(\ref{ccl}): using again relation (\ref{ax}), after integration, one ends up with the more accurate expression~:

\begin{equation}\label{al} \Th_\pm(t) = \pm \al - A_x^0 \ca t^2 - \frac{1}{3} \sinca \ca t^3 +O(t^4). \end{equation}

Following the same reasoning, we shall now consider particles initially positioned in correspondence of the upper, alternatively lower, boundary of the 
rectangular water--bag profile, $(\theta(0)=0,p(0)=\pm \Dp/2)$. These particles are also virtually linked to the boundary that they contribute to create and thus 
$p(t)=P_\pm(\theta(t),t)$. Recalling that the phases evolve as $\theta(t)=\pm t\Dp /2 + O(t^2)$, one can solve  
Eq.(\ref{hampart}) to obtain the following expression for the conjugate momenta~:

\begin{equation} P_\pm(\theta,t) = \pm \frac{\Dp}{2} -2 A_x^0 t -\sinca t^2 +O(t^3) \end{equation}

Then, as that $P_\pm= u \theta^2 + v_\pm$, and since $u(t)$ goes as $O(t)$, one can conclude that~:

\begin{equation}\label{vpm} v_\pm = \pm \frac{\Dp}{2} - 2 A_x^0 t - \sinca t^2 + O(t^3). \end{equation}

The conservation of the total momentum which, in the Vlasov picture, reads $P=A_x^2+A_y^2 + \int \! \! \int f(\theta,p,t) p\ d\theta\ dp$, now takes the form~:

\begin{equation}  I_0 = A_x^2+A_y^2 + f_0 (\frac{1}{3}(\Th_+^3-\Th_-^3) u \Dv + \frac{1}{2}(\Th_+ - \Th_-) \Sv \Dv), \label{eqmom}\end{equation}

where $\Sv=v_+ + v_-$ and $\Dv=v_+ - v_-$ have been introduced. Moreover from Eqs.~(\ref{ax}) and (\ref{ay}) that~:

\begin{equation} A_x^2+A_y^2 = I^0 + 2 A_x^0 \sinca t + \sinca^2 t^2 + O(t^3).
\end{equation}

Expression (\ref{eqmom}) can be hence manipulated by 
making also use of expansions (\ref{al}) and (\ref{vpm}). Solving for $u(t)$ results in~:
\begin{equation}\label{eq:u} u(t)= \frac{3}{2 \al^2} (\sinca -1) \Sv + O(t^3) = \frac{6}{\al^2}(1-\sinca)A_x^0 t + \frac{3}{\al^2} \sinca (1 - \sinca) t^2 + O(t^3). \end{equation}

Furthermore, the conservation of the energy reads:
\begin{equation} H=\int \! \! \int f(\theta,p,t) \frac{p^2}{2}\ d\theta\ dp + 2 \int \! \! \int f(\theta,p,t) (A_x \sin{\theta} +A_y  \cos{\theta}) d\theta\ dp,\end{equation}
which leads to~:
\begin{eqnarray} \frac{\Dp^2}{24} &=&\frac{f_0}{6} [ \frac{3}{5}(\Th_+^5 - \Th_-^5) u^2 \Dv + (\Th_+^3 - \Th_-^3) u \Sv \Dv \\
&+&(\Th_+ - \Th_-) \frac{\Dv}{4}(\Dv^2+3\Sv^2) ] + 2 (A_y \dot{A_x} - A_x \dot{A_y})
\nonumber. 
\end{eqnarray}

The l.h.t. is the initial energy of the system: since the initial water-bag is centered around zero, the field does not contribute to the
energy at $t=0$ and the only residual component comes from particles' kinetic energy. 
Finally, replacing in the above expression each term - apart from $A_y$ - by its expansion (see equations (\ref{al}), (\ref{vpm}) and (\ref{eq:u})), we get for $A_y$ a first-order differential equation. Then, assuming the general form $A_y(t)= \eta t^n + \nu t^{n+1} +O(t^{n+2})$, one immediately realizes that 
$n=3$, and, more precisely~:
\begin{equation}\label{ay2} A_y(t) = \frac{A_x^0}{15}(4 - 8 \sinca + 9 \sinca^2)  t^3 + \frac{\sinca}{60}(4 - 8 \sinca + 9 \sinca^2) t^4 + O(t^5). \end{equation}

Note that the expansions here derived can be used as a starting point to calculate higher order corrections, following a typical strategy often employed in
 perturbative analysis. In particular, from Eqs.(\ref{dax}), (\ref{al}) and (\ref{vpm}), we get~:
\begin{equation} A_x(t)=A_x^0 + \sinca t +O(t^4). \end{equation}

It should be however stressed that this procedure cannot converge indefinitely, since the assumptions built into the model will eventually prove inaccurate and 
further effects will need to be properly incorporated (e.g. the leaning of the lateral boundaries).

\section{Predicting the macroscopic observables}\label{macroscopic}

In the above paragraph we have developed a perturbative approach that ultimately enabled us to provide closed 
analytic expressions for the time evolution of the complex field $\bm{A}$ and the single--particle distribution function $f(\theta,p,t)$.
This novel insight allows us to condensate in compact formulae the time dependence of all fundamental macroscopic quantities, 
modified during the self-consistent amplification process. Few examples are discussed in the remaing part of this Section.

The wave intensity $I(t)$ follows trivially as:

\begin{equation}\label{intensity} I(t) = |\bm{A}|^2 = A_x(t)^2 +A_y(t)^2 = I_0 + 2\sqrt{I_0} \sinca t + \sinca^2 t^2 +O(t^4). \end{equation}

\begin{figure}[ht]
\centerline{
\epsfig{figure=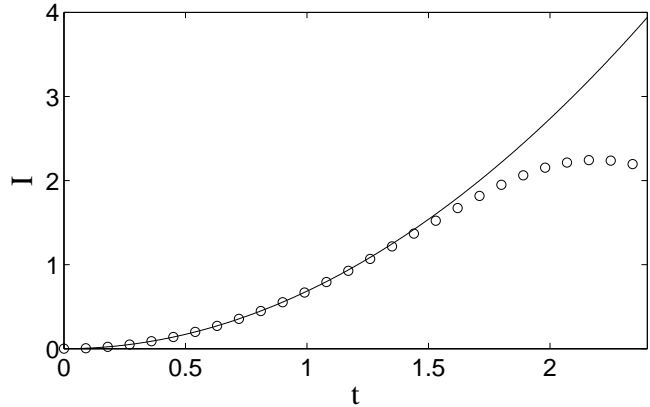,width=10cm}}
\caption{Time evolution of the laser intensity. Here $I_0 \simeq 0$. Particles are initially distributed in the interval $[-\frac{\pi}{3};\frac{\pi}{3}]$. The solid line refers to the quadratic law predicted by Eq.(\ref{intensity}). The circles represents the numerical simulations based on Hamiltonian~(\ref{eq:Hamiltonian}).\label{fig2}}
\end{figure}

For $I_0=0$, the laser intensity scales quadratically with time, a result previosuly reported in \cite{Yu}. This finding agrees with direct numerical simulations based on the 
N-body model (\ref{eq:Hamiltonian})
and reported in figure \ref{fig2}. Clearly, our solution is limited to non-homogeneous (bunched) initial beam: If
the phases of the particles are initially occupying the whole interval $[-\pi,\pi]$, an exponential instability develops, as predicted by the linear analysis of
section \ref{model}.

\begin{figure}[ht]
\centerline{
\epsfig{figure=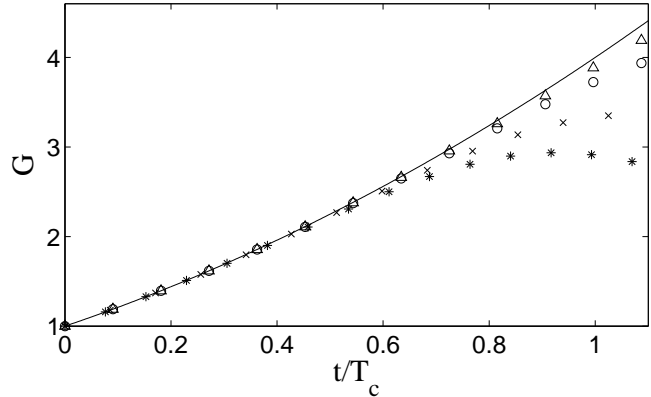,width=10cm}}
\caption{The gain $G$ is plotted as function of the rescaled time $t/T_c$. The plain line refers to the theoretical prediction (\ref{gain}), the symbols to numerical simulations~: the crosses correspond to $I_0=0.8N$ and $\al=\pi/2$, the circles to $I_0=0.8N$and $\al=\pi/4$, the triangles for $I_0=0.4N$ and $\al=\pi/2$, while the stars stand for $I_0=N$ and $\al=\pi/2$. We shall here notice that equation 
(\ref{gain}) is found to be accurate for initial bunching and intensity resp. smaller than $\pi/2$ and $0.8N$. For larger values, higher order corrections need to be incorporated into the model.\label{fig3}}
\end{figure}

Consider now the case $I_0 \ne 0$ and define the gain $G(t) = I(t)/I_0$. One can hence recast equation (\ref{intensity}) in the form~:

\begin{equation} \label{gain} G(t) = 1 + 2 \frac{t}{T_c} + \left(\frac{t}{T_c}\right)^2+O\left(\left(\frac{t}{T_c}\right)^3\right). \end{equation}

where we have introduced the characteristic time $T_c=\sqrt{I_0}/s_\al$. In principle, increasing $T_c$ amounts to slow down the growth: 
longer times (undulators) are thus required to attain a fixed gain level. Interestingly, a faster evolution is produced when increasing the initial 
particle bunching. A similar effect is obtained by reducing the intensity $I_0$ of the injected seed. In figure  \ref{fig3}, we report the gain $G$ 
as funtion of the rescaled time $t/T_c$: symbols refer to numerical simulations based on Hamiltonian (\ref{eq:Hamiltonian}). The data are nicely interpolated
by the universal profile (\ref{gain}). By inverting equation (\ref{gain}), one can estimate for time $t^*$  needed to the system to
reach a fixed gain amount $G^*$. A straightforward calculation leads to the following compact relation

\begin{equation} 
t^* =  \frac{\sqrt{I_0}}{s_\al} \left( \sqrt{G^*} -1 \right) ,
\end{equation}

which can be used as a first rough guideline for optimization and design purposes. 

As a second example, we consider the particles' energy dispersion here defined as~:


\begin{equation}\label{dispersion} D = \int \! \! \int d\theta\ dp\ f\ p^2 - (\int \! \! \int d\theta\ dp\ f\ p)^2. \end{equation}

From the above, after some algebra, it follows~:

\begin{equation} \label{dispersion1}  D(t) = \frac{\Dp^2}{12} + \frac{16}{5} I_0^2 \frac{(\sinca-1)^2}{\sinca^2} \left[  \left(\frac{t}{T_c} \right )^2+\left(\frac{t}{T_c}\right)^3
+O\left(\left(\frac{t}{T_c}\right)^4\right) \right]. \end{equation}

For small values of $I_0$, the wave intensity increases as $\sinca^2 t^2$ and hence   the 
particles' energy scatters as $t^4$~:

\begin{equation} D(t) = \frac{\Dp^2}{12} + \frac{1}{5} (4 \sinca^4 - 8 \sinca^3 + 4 \sinca^2) t^4 + O(t^5). \end{equation}

\begin{figure}[ht]
\centerline{
\epsfig{figure=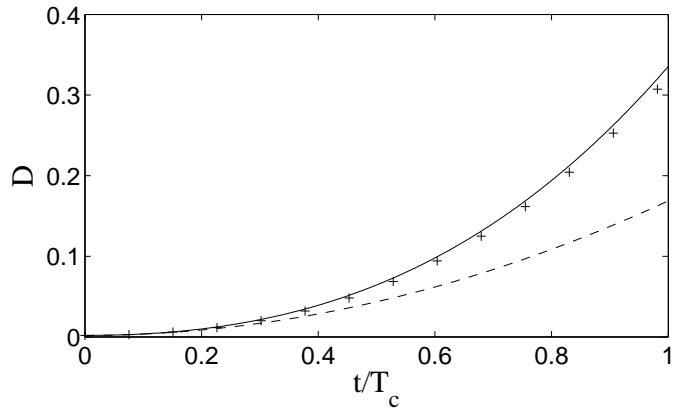,width=10cm}}
\caption{Diffusion $D(t)$ vs time $t$. Symbols refers to the N-body simulation, while the dashed and solid lines represents the theoretical prediction~(\ref{dispersion1})~: the former stands for the second order in $t$ prediction, the latter for the third order.\label{fig4}   
}
\end{figure}

Again, the theory agrees well with direct simulations as reported in Fig.\ref{fig4} (b).  

Finally, the bunching coefficients (\ref{bunch}) read~:

\begin{equation} b_k(t) = \frac{\sin{(k\al)}}{k\al} + O(t^3). \qquad for \quad k=1,2..  \end{equation}

This in turn implies that there is no significant change in the bunching of the particles during the early stage of the interaction.

\begin{figure}[ht]
\centerline{
\epsfig{figure=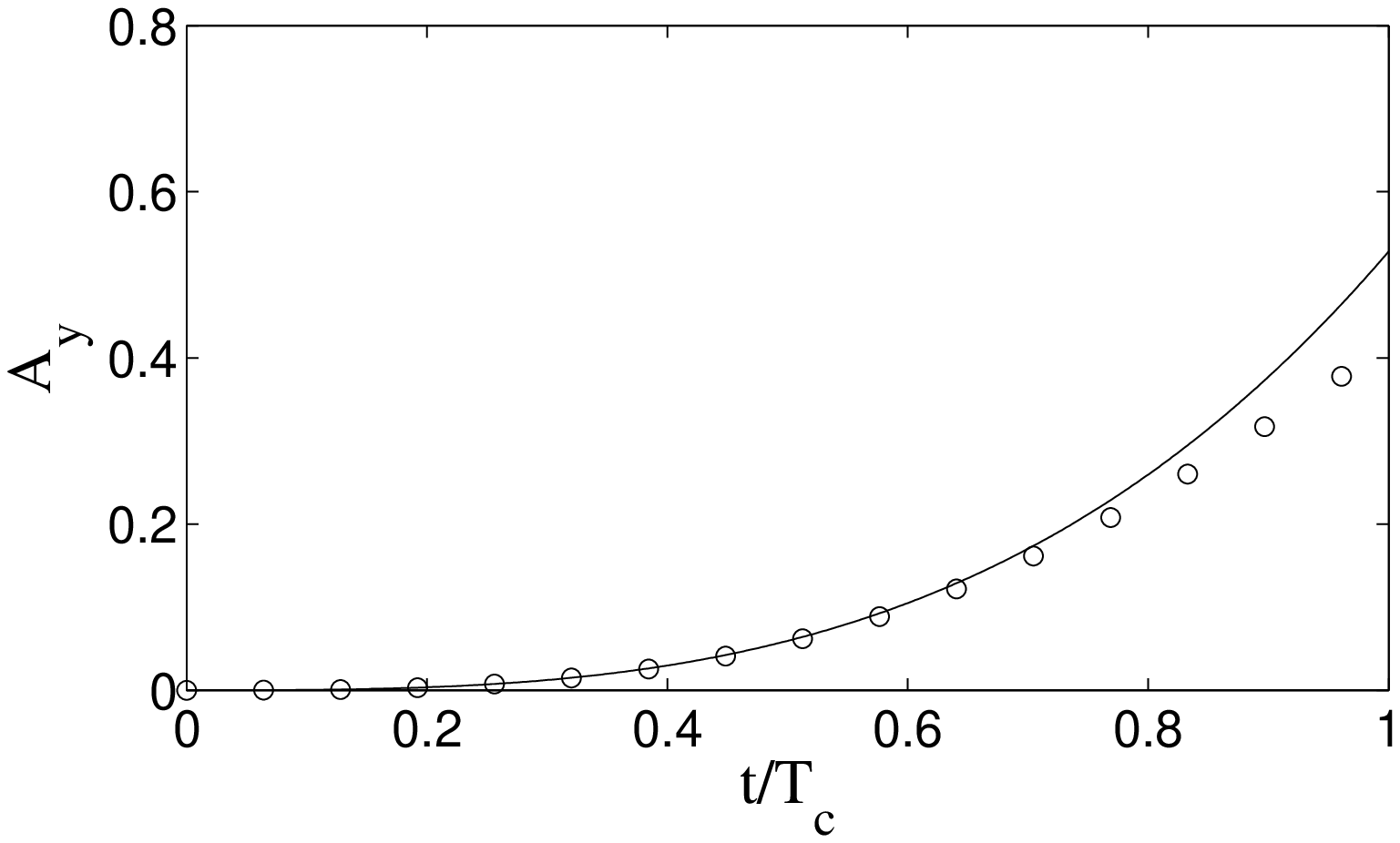,width=10cm}}
\caption{$A_y(t)$ vs time $t$. Symbols refers to the N-body simulation, while the solid line represents the theoretical prediction~(\ref{ay2}).\label{fig5}   
}
\end{figure}

As a side remark we shall notice that the higher order corrections  
for $A_y$ as predicted by equation (\ref{ay2}) correlate very well with  
the numerics, see Fig.~\ref{fig5}.

\section{Conclusion}\label{conclusion}

In this paper we developed a perturbative approach to characterize the short-time evolution of 
a Single-Pass Free Electron Laser. In particular we provide closed analytical expressions 
that elucidate the time dependence of the main macroscopic quantities, e.g. laser intensity, degree of bunching and energy 
dispersion. More specifically, particles are initially randomly distributed inside in-homogeneous (spatially bunched) water-bag envelopes. The  
underlying Vlasov dynamics results in a progressive distorsion of the water-bag profile, which we here monitor by tracking the  
evolution of the lateral boundaries. The prediction of the theory 
are shown to agree with direct numerical simulations. Interestingly, we also derive a universal relation which 
allows one to calculate the laser intensity gain at a given undulator length.  

In conclusion, it is worth emphasising that we here address the study of a generic 
wave-particles interaction process. It can be thereofore expected that our conclusions will 
prove useful beyond the realm of FEL applications, and possibly translate to other fields 
where the complex interplay between particles and waves is known to be crucial.

\end{document}